# Progress on a Conjecture Regarding the Triangular Distribution

Hien D. Nguyen[1,2] and Geoffrey J. McLachlan[1]

November 5, 2016

[1]School of Mathematics and Physics, University of Queensland. [2]Centre for Advanced Imaging, University of Queensland.

## Abstract

Triangular distributions are a well-known class of distributions that are often used as an elementary example of a probability model. Maximum likelihood estimation of the mode parameter of the triangular distribution over the unit interval can be performed via an order statistic-based method. It had been conjectured that such a method can be conducted using only a constant number of likelihood function evaluations, on average, as the sample size becomes large. We prove two theorems that validate this conjecture. Graphical and numerical results are presented to supplement our proofs.

## 1 Introduction

Let $X \in [0, 1]$ be a random variable with cumulative distribution function (CDF)

$$F_\theta(x) = \begin{cases} x^2/\theta, & \text{if } x < \theta, \\ 1 - (1-x)^2/(1-\theta), & \text{if } x \geq \theta, \end{cases} \quad (1)$$

when $\theta \in (0,1)$, $F_0(x) = 1 - (1-x)^2$ when $\theta = 0$, and $F_1(x) = x^2$ when $\theta = 1$. We say that $X$ arises from a triangular distribution with mode parameter $\theta \in [0,1]$. The probability density function of $X$ can be written as

$$f_\theta(x) = \begin{cases} 2x/\theta, & \text{if } x < \theta, \\ 2(1-x)/(1-\theta), & \text{if } x \geq \theta, \end{cases}$$

when $\theta \in (0,1)$, $f_0(x) = 2(1-x)$ when $\theta = 0$, and $f_1(x) = 2x$ when $\theta = 1$.

The triangular distribution is a popular probability model for teaching, due to its simple geometric form; see Doane (2004) and Price and Zhang (2007) for examples where the triangular distribution is used in the teaching of various aspects of distribution theory. Outside of the classroom, the triangular distribution has also been used to model task completion times for Program Evaluation and Review Techniques (PERT) models, prices of securities that are traded on the New York Stock Exchange, and haul times in civil engineering data. Elaborations on these applications can be found in Kotz and Van Dorp (2004, Ch. 1) and the references therein.

Recently, there has been a renewed interest in the triangular distribution. For example, Glickman and Xu (2008) investigated the distribution of the product of triangular distributions for applications in traffic-related risk assessment, Karlis and Xekalaki (2008) considered the use of mixtures of triangular distributions for estimation of bounded and concave densities, and Nagaraja (2013) derived expressions for the moments of order statistics and L-moments, for application smart communication networks. Furthermore, Gunduz and Genc (2015) followed the work of Glickman and Xu (2008) and derived expressions for the distribution of quotients of triangular distributions, and Nguyen and McLachlan (2016) utilized a novel characterization of the triangular distribution to derive minorization–maximization algorithms (Hunter and Lange, 2004) for the maximum likelihood (ML) estimation of the triangular distribution and the mixture of triangular distributions of Karlis and Xekalaki (2008).

Let $X_1, ..., X_n$ be a random IID (independent and identically distributed) sample of size $n \in \mathbb{N}$ from a triangular distribution with unknown mode parameter and let $x_1, ..., x_n$ be its realization. The likelihood function and the ML estimator can be expressed as $\mathcal{L}_n(\theta) = \prod_{i=1}^{n} f_\theta(x_i)$ and

$$\hat{\theta}_n = \arg\max_{\theta \in [0,1]} \mathcal{L}_n(\theta),$$

respectively.

Let $x_{(1)} \leq x_{(2)} \leq ... \leq x_{(n)}$ be the order statistics of the sample realization. It is shown in Oliver (1972) that $\hat{\theta}_n = x_{(i)}$ for some $i = 1, ..., n$. That is, the ML estimate is always one of the observations from sample; the same result was proved in Kotz and Van Dorp (2004, Ch. 1). Remarkably, Oliver (1972) also showed that if $\hat{\theta}_n = x_{(j)}$, then $x_{(j)}$ must fulfill the condition $(j-1)/n < x_{(j)} < j/n$. Thus, the ML estimator can be rewritten as

$$\hat{\theta}_n = \arg\max_{\theta \in \Theta_n} \mathcal{L}_n(\theta) \qquad (2)$$

where $\Theta_n = \{x_{(j)} : (j-1)/n < x_{(j)} < j/n, \ j = 1, ..., n\}$.

An investigation into the expected number of elements in the set $\Theta_n$ was conducted by Huang and Shen (2007). Let $m_n = \sum_{j=1}^{n} \mathbb{I}\{(j-1)/n < x_{(j)} < j/n\}$ be the observed number of elements in $\Theta_n$, where $\mathbb{I}\{A\}$ equals 1 if proposition $A$ is true and 0 otherwise. As with Huang and Shen (2007), we will subsequently call $m_n$ the observed number of 'matches'. It was observed and conjectured that

$$\mathbb{E}(M_n) = \sum_{j=1}^{n} \mathbb{P}\left(\frac{j-1}{n} < X_{(j)} < \frac{j}{n}\right) \approx 2 \qquad (3)$$

as $n$ grows large, for various values of $\theta$. Here $X_{(1)} \leq X_{(2)} \leq ... \leq X_{(n)}$ is the order statistics of a random sample, and $M_n = \sum_{j=1}^{n} \mathbb{I}\{(j-1)/n < X_{(j)} < j/n\}$.

In contrast to this observation regarding the triangular distribution, Huang and Shen (2007) proved that if one assumes that $X_1, ..., X_n$ is an IID sample from a uniform distribution over the unit interval (i.e. $X \in [0, 1]$), then the expected number of elements in $\Theta_n$ is

$$\mathbb{E}(M_n) = 1 + \sum_{j=1}^{n-1} \binom{n}{j} \left(\frac{j}{n}\right)^j \left(1 - \frac{j}{n}\right)^{n-j}. \qquad (4)$$

Upon application of Stirling's formula [cf. Charalambides (2002, Thm. 3.2)] and $\lim_{n \to \infty} \sum_{j=1}^{n-1} 1/\sqrt{j(n-j)} = \pi$, Huang and Shen (2007) obtained the approximation formula

$$\mathbb{E}(M_n) \approx 1 + \sqrt{\frac{n}{2\pi}} \sum_{j=1}^{n-1} \frac{1}{\sqrt{j(n-j)}} \approx 1 + \sqrt{\frac{\pi n}{2}} \qquad (5)$$

for large $n$. Thus, whereas the expected number of matches for a triangular distribution remains constant as $n$ grows, the expected number of matches grows at a rate of $O(\sqrt{n})$ [using Landau's O-notation; see Cormen et al. (2002, Ch. 3)] asymptotically, when the sample arises from a uniform distribution.

In Nguyen and McLachlan (2016), it was established that the average order of complexity for computing the ML estimate using the estimator (2) is $O(n[\log n + \mathbb{E}(M_n)])$. This is due to the requirement of a sorting algorithm to obtain the order statistics, which requires $O(n \log n)$ operations [e.g. Heapsort (Cormen et al., 2002, Ch. 6, p. 146)], and the computation of the likelihood function for comparison, which requires $O(n)$ operations. Thus, if the conjecture of Huang and Shen (2007) is true (i.e. (3) is true), then the ML estimate has average order of complexity $O(n \log n)$, which makes it no more complex than the computation of order statistics.

In this article, we extend the work of Huang and Shen (2007) to obtain an expression for the expected number of matches $\mathbb{E}(M_n)$ of samples arising from any distribution over the unit interval. Furthermore, we prove that for any continuous distribution, the maximum rate of growth of $\mathbb{E}(M_n)$ is $O(\sqrt{n})$.

Using the derived general expression for $\mathbb{E}(M_n)$, we provide a formula for the expected number of matches of any sample arising from a triangular distribution. Using the formula, we then prove that $\mathbb{E}(M_n) \to 1.684567$, as $n \to \infty$, for $\theta \in \{0, 1\}$. Similarly, for $\theta \in (0, 1)$, we obtain the approximation $\mathbb{E}(M_n) \approx N_n$ for large $n$, where $N_n \to 2.369134$, as $n \to \infty$. Thus, we obtain positive progress towards the conjecture of Huang and Shen (2007). We supplement the main result with graphical and numerical results regarding the expected number of matches from a triangular distributed sample.

Although the content of the article is aimed at resolving the discussed conjecture of Huang and Shen (2007), it is anticipated that the techniques from this article can also be applied in other settings. The standard two-sided power distribution (STPD) of Kotz and Van Dorp (2004, Ch. 3) is a generalization of the triangular distribution for which the ML estimator is an order statistic that satisfies a criterion [cf. Kotz and Van Dorp (2004, Ch. 3, p. 80)]. It may be possible to apply the methodology from this article to obtain an analogous result in the STPD setting. The beta distribution is also interesting as it includes both the $\theta \in \{0, 1\}$ cases of the triangle as well as the uniform distribution as special cases. An interesting problem that arises is to determine the parameter settings for which the number of matches $\mathbb{E}(M_n)$ converges in the beta distribution setting.

The rest of the article proceeds as follows. General results regarding the expected number of matches that are obtained from arbitrary distributions over the unit interval are presented in Section 2. Specific results regarding the triangular distribution are presented in Section 3. Numerical and graphical results are presented in Section 4. Proofs for the main results are relegated to Section 5.

## 2 General Results

Let $X_1, ..., X_n$ be an IID random sample, where $X_1 \in [0,1]$ has distribution function $F(x)$, which is well-defined for $x \in [0,1]$. Let, $X_{(1)} \leq X_{(2)} \leq ... \leq X_{(n)}$ be the order statistics of the random sample. For $j = 1, ..., n$, let $F_{j:n}(x)$ be the distribution function of $X_{(j)}$. David and Nagaraja (2003, Eq. 2.1.3) states

$$F_{j:n}(x) = \sum_{i=j}^{n} \binom{n}{i} F^i(x) [1 - F(x)]^{n-i}, \qquad (6)$$

which provides the link between $F(x)$ and the order statistic distributions. Using (6), we get the general formula for the expected number of matches of any sample arising from distribution $F(x)$.

**Theorem 1.** *Let $X_1, ..., X_n$ be an IID random sample, where $X_1 \in [0,1]$ has distribution $F(x)$. The expected number of matches from the sample is*

$$\mathbb{E}(M_n) = 1 + \sum_{j=1}^{n-1} \binom{n}{j} F^j\left(\frac{j}{n}\right) \left[1 - F\left(\frac{j}{n}\right)\right]^{n-j}. \qquad (7)$$

*Remark* 1. If $X_1$ has distribution $F(x) = x$ (i.e. $X_1$ is uniformly distributed), then (7) becomes (4), as expected.

Via an elementary calculus argument and upon application of Stirling's for-mula, we obtain the following upper bound for the asymptotic growth rate of the expected number of matches.

**Corollary 1.** *Let $X_1, ..., X_n$ be an IID random sample, where $X_1 \in [0, 1]$ has distribution $F(x)$. For any $F(x)$, the asymptotic growth rate of $\mathbb{E}(M_n)$ has upper bound $O(\sqrt{n})$. For large $n$, the upper bound of $\mathbb{E}(M_n)$ can be asymptotically approximated by Equation (5).*

*Remark* 2. Corollary 1 implies that the uniform distribution maximizes the growth rate of the expected number of matches. Interestingly, the right-hand side of (5) is related to the enumeration of rooted trees by total height. That is, $n\sqrt{\pi n/2}$ is the asymptotic growth rate of the mean total height of all rooted trees with $n$ labeled points, and also the mean height; see Riordan and Sloane (1969). Furthermore, upon rearrangement of Equation (4), we obtain the OEIS (On-line Encyclopedia of Integer Sequences; https://oeis.org) sequence A001864 from the expression $n^n (\mathbb{E}(M_n) - 1)$; see also M2138 from Sloane and Plouffe (1995).

## 3 Triangular Distribution

Let $X_1, ..., X_n$ be an IID random sample, where $X_1$ has triangular distribution function $F_\theta(x)$. From Theorem 1, we have

$$\mathbb{E}(M_n) = 1 + \sum_{j=1}^{n-1} \binom{n}{j} F_\theta^j \left(\frac{j}{n}\right) \left[1 - F_\theta \left(\frac{j}{n}\right)\right]^{n-j}. \tag{8}$$

In the case where $\theta = 0$ or $\theta = 1$, (8) becomes

$$\mathbb{E}(M_n) = 1 + \sum_{j=1}^{n-1} \binom{n}{j} \left[1 - \left(1 - \frac{j}{n}\right)^2\right]^j \left(1 - \frac{j}{n}\right)^{2(n-j)}$$

and

$$
\begin{aligned}
\mathbb{E}(M_n) &= 1 + \sum_{j=1}^{n-1} \binom{n}{j} \left(\frac{j}{n}\right)^{2j} \left[1 - \left(\frac{j}{n}\right)^2\right]^{n-j} \\
&= 1 + \sum_{j=1}^{n-1} \binom{n}{n-j} \left(\frac{n-j}{n}\right)^{2(n-j)} \left[1 - \left(\frac{n-j}{n}\right)^2\right]^{j} \\
&= 1 + \sum_{j=1}^{n-1} \binom{n}{j} \left[1 - \left(1 - \frac{j}{n}\right)^2\right]^{j} \left(1 - \frac{j}{n}\right)^{2(n-j)},
\end{aligned}
\tag{9}
$$

respectively. It is not difficult to obtain the limit

$$\lim_{n \to \infty} \binom{n}{j} \left[1 - \left(1 - \frac{j}{n}\right)^2\right]^{j} \left(1 - \frac{j}{n}\right)^{2(n-j)} = \frac{2^j e^{-2j} j^j}{j!} = s_j.$$

Thus, we have

$$\mathbb{E}(M_n) \approx 1 + S_{n-1} \tag{10}$$

for large $n$ in the cases where $\theta \in \{0, 1\}$, and $S_n = \sum_{j=1}^{n} s_j$.

**Lemma 1.** *The series $S_n$ converges to $S_\infty \approx 0.684567$.*

Applying Lemma 1 to (10) yields the first main result of the article.

**Theorem 2.** *For triangular distributions with $\theta \in \{0, 1\}$, $\mathbb{E}(M_n) \to 1.684567$, as $n \to \infty$.*

Suppose now that $n = p + q$ and let $\theta = p/(p+q)$, where $p, q \in \mathbb{N}$. From (1), we note that $F_\theta(\theta) = \theta$; thus $\theta$ is the $\theta$th quantile of the distribution. Using this fact, we have the following expression of (8):

$$
\begin{aligned}
\mathbb{E}\left(M_{p+q}\right) &= 1+\sum_{j=1}^{p}\binom{p+q}{j} F_\theta^j\left(\frac{j}{p+q}\right)\left[1-F_\theta\left(\frac{j}{p+q}\right)\right]^{p+q-j} \\
&\quad+\sum_{j=p+1}^{p+q-1}\binom{p+q}{j} F_\theta^j\left(\frac{j}{p+q}\right)\left[1-F_\theta\left(\frac{j}{p+q}\right)\right]^{p+q-j} \\
&= 1+\sum_{j=1}^{p}\binom{p+q}{j} F_1^j\left(\frac{j}{p}\right)\left[1-F_1\left(\frac{j}{p}\right)\right]^{p+q-j} \\
&\quad+\sum_{j=p+1}^{p+q-1}\binom{p+q}{j} F_0^j\left(\frac{j}{q}\right)\left[1-F_0\left(\frac{j}{q}\right)\right]^{p+q-j} \\
&= 1+\sum_{j=0}^{p}\binom{p+q}{p-j}\left(\frac{p-j}{p}\right)^{2(p-j)}\left[1-\left(\frac{p-j}{p}\right)^2\right]^{q+j} \\
&\quad+\sum_{j=1}^{q-1}\binom{p+q}{p+j}\left[1-\left(1-\frac{p+j}{q}\right)^2\right]^{p+j}\left(1-\frac{p+j}{q}\right)^{2(q-j)}.
\end{aligned}
$$

The second equality is due to the fact that the first $\theta = p/(p+q)$ is the $[p/(p+q)]$ th quantile. We apply this fact by noting that $f_\theta(x)$, to the left of $\theta$, has upwards-sloping density and vice versa, has downward-sloping density to the right. Since all triangular distributions must have height $f_\theta(\theta) = 2$, the region to the left of $\theta$ is proportional $f_1(x)$ in the sense that $f_\theta(x) = f_1(y)$, for $x = \theta y$, and similarly the region to the right is proportion to $f_0(x)$ in the sense that $f_\theta(x) = f_0(y)$, for $x = (1-\theta)y$. Via the proportionality argument, we can compute

$$F_\theta\left(\frac{j}{n}\right) = F_1\left(\frac{j}{n}\times\frac{n}{p}\right) = F_1\left(\frac{j}{p}\right),$$

for $j/n \leq \theta$, and

$$F_\theta\left(\frac{j}{n}\right) = F_0\left(\frac{j}{n}\times\frac{n}{q}\right) = F_0\left(\frac{j}{q}\right),$$

for $j/n > \theta$.

Again, it is not difficult to obtain the limits

$$\lim_{p\to\infty} \binom{p+q}{p-j} \left(\frac{p-j}{p}\right)^{2(p-j)} \left[1-\left(\frac{p-j}{p}\right)^2\right]^{q+j} = \frac{2^{j+q}e^{-2j}j^{j+q}}{(j+q)!} = u_j^q$$

and

$$\lim_{q\to\infty} \binom{p+q}{p+j} \left[1-\left(1-\frac{p+j}{q}\right)^2\right]^{p+j} \left(1-\frac{p+j}{q}\right)^{2(q-j)} = \frac{2^{j+p}e^{-2(j+p)}(j+p)^{j+p}}{(j+p)!} = v_j^p.$$

Thus, we have

$$\mathbb{E}(M_{p+q}) \approx 1 + U_p^q + V_q^p, \tag{11}$$

where $U_p^q = \sum_{j=1}^p u_j^q$ and $V_q^p = \sum_{j=1}^q v_j^p$, for large $p$ and $q$. Note that $U_p^q$ is indexed from $j=1$, since $0^{0+q} = 0$ for any $q \in \mathbb{N}$.

**Lemma 2.** *The series $U_p^q$ converges as $p \to \infty$, for any finite $q \in \mathbb{N}$, and the series $V_q^p$ converges as $q \to \infty$, for any finite $p \in \mathbb{N}$.*

For fixed $j$, we note that $u_j^q/s_j = 2^q j^q j!/(j+q)! \to 0$, as $q \to \infty$ since factorials grow faster than exponentials. Similarly $v_j^p/s_j \to 0$, as $p \to \infty$. Thus, we obtain the following result.

**Lemma 3.** *The ratios $u_j^q/s_j \to 0$ and $v_j^p/s_j \to 0$, as $q \to \infty$ and $p \to \infty$, respectively, for any finite $j \in \mathbb{N}$.*

Lemmas 2 and 3 suggest that for large $p$ and $q$, $\mathbb{E}(M_{p+q})$ converges to a finite constant. Further, it is suggested that we can conservatively approximate $\mathbb{E}(M_{p+q})$ by $N_{p,q} = 1 + S_p + S_q$, in the sense that $1 + U_p^q + V_q^p \leq N_{p,q}$ for large $p$ and $q$. An application of Lemma 1 yields the second main result of the article.

**Theorem 3.** *For triangular distributions with $\theta = p/(p+q)$ for $p, q \in \mathbb{N}$, $\mathbb{E}(M_{p+q}) \approx N_{p,q}$, where $N_{p,q} \to 2.369134$, as $p \to \infty$ and $q \to \infty$.*

Table 1: Exact values of expected matches $\mathbb{E}(M_n)$ for various sample sizes $n$ and mode parameters $\theta$.

| $n\backslash\theta$ | 0 | 0.1 | 0.2 | 0.3 | 0.4 | 0.5 |
|---:|---:|---:|---:|---:|---:|---:|
| 5 | 1.814694 | 1.960713 | 2.093680 | 2.162657 | 2.207036 | 2.217071 |
| 10 | 1.929234 | 2.271707 | 2.466275 | 2.579562 | 2.640312 | 2.659541 |
| 20 | 1.842068 | 2.394180 | 2.608803 | 2.707651 | 2.751632 | 2.764140 |
| 50 | 1.731881 | 2.490839 | 2.531784 | 2.487879 | 2.450077 | 2.436588 |
| 100 | 1.705885 | 2.477835 | 2.358059 | 2.282239 | 2.249001 | 2.239780 |
| 200 | 1.694797 | 2.332438 | 2.205683 | 2.166804 | 2.151832 | 2.147659 |
| 500 | 1.688567 | 2.162322 | 2.110494 | 2.093518 | 2.086342 | 2.084280 |
| 1000 | 1.686553 | 2.103114 | 2.073165 | 2.062607 | 2.058058 | 2.056742 |

## 4 Graphical and Numerical Results

Using **R** (R Core Team, 2013), we compute exact values of expected matches $\mathbb{E}(M_n)$ for values of $n \leq 1000$, and for triangular distributions with parameters $\theta \in \{0, 0.1, 0.2, 0.3, 0.4, 0.5\}$. The **ptriangle** function from the package **triangle** (Carnell, 2016) and the **chooseZ** function from the package **gmp** (Lucas et al., 2014) were used to evaluate the CDF of the triangular distribution and precisely evaluate the necessary combinatorial values, respectively.

Figure 1 indicates that for $\theta \in \{0, 1\}$, the approximated limit of 1.684567 appears sharp (note that $\mathbb{E}(M_n)$ is symmetric in $\theta$ about $1/2$). However, for $\theta \in (0, 1)$, the approximated limit of 2.369134 appears quite conservative, as remarked in Section 3. We also observe that for small values of $n < 200$, the behavior of $\mathbb{E}(M_n)$ lacks monotonicity; however, when $n \geq 200$, $\mathbb{E}(M_n)$ appears to be decreasing for all cases of $\theta$. Table 1 extends upon Huang and Shen (2007) by tabulating the exact values of $\mathbb{E}(M_n)$ for $n = 1000$ and also increasing the precision from 4 to 6 decimal places.

To supplement the results from Figure 1 and Table 1, which replicate the study by Huang and Shen (2007) with greater accuracy, we also consider the $\theta \in \{0.05, 0.15, 0.25, 0.35, 0.45\}$ cases. The computations follow the same setup and the tabulation and graphical representation of the results are provided in

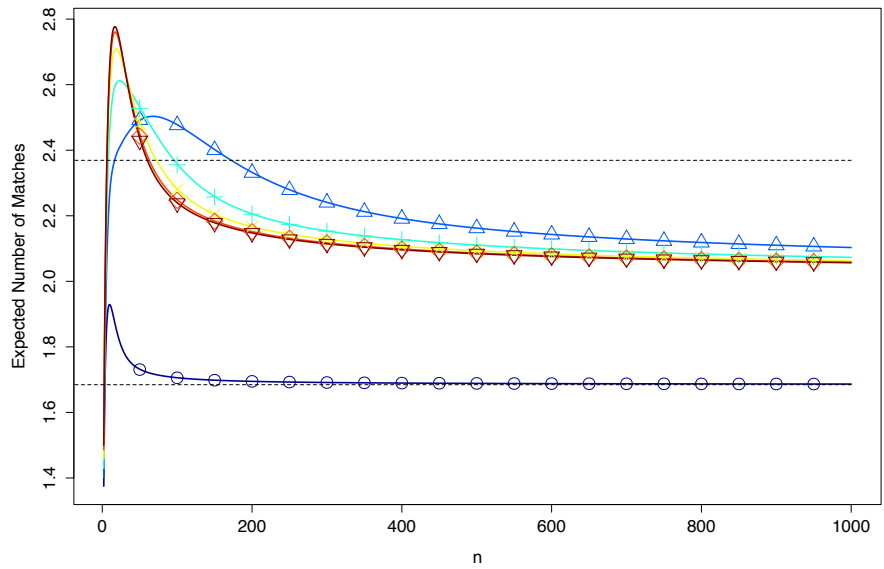

Figure 1: The lines marked by the symbols ○, △, +, ×, ⋄, and ∇ indicate the exact values of expected matches $\mathbb{E}(M_n)$ for triangular distributions with $\theta$ set to 0, 0.1, 0.2, 0.3, 0.4, and 0.5, respectively. The lower and upper dashed lines mark the values 1.684567 and 2.369134, respectively.

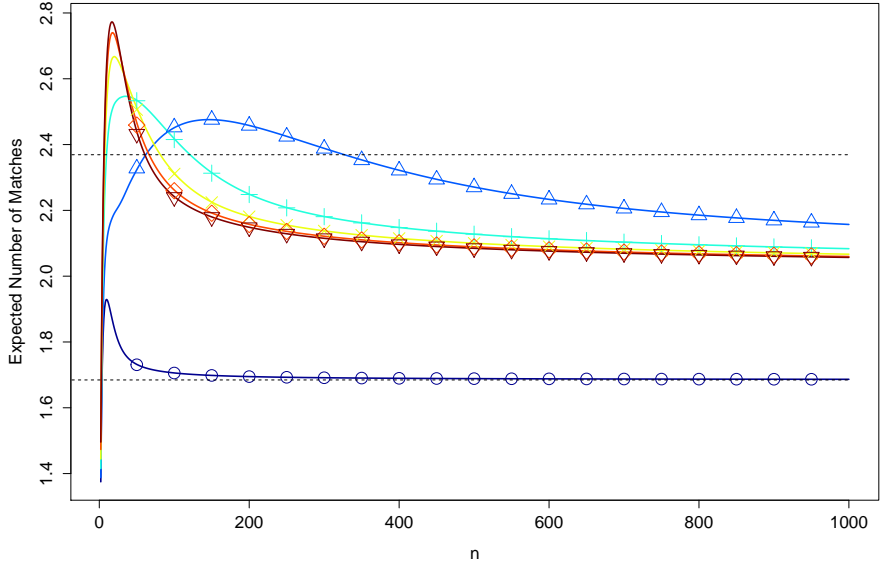

Figure 2: The lines marked by the symbols ∘, △, +, ×, ⋄, and ∇ indicate the exact values of expected matches $\mathbb{E}(M_n)$ for triangular distributions with $\theta$ set to 0, 0.05, 0.15, 0.25, 0.35, and 0.45, respectively. The lower and upper dashed lines mark the values 1.684567 and 2.369134, respectively.

Table 2 and Figure 2, respectively. Like the $\theta \neq 0$ cases that are considered by Huang and Shen (2007), we also observe that the supplementary cases appear to exhibit convergence to some value below the approximated limit of 2.369134. Furthermore, the behavior of $\mathbb{E}(M_n)$ also lacks monotonicity for small values of $n$.

Using 1000 Monte Carlo simulations of $n \in \{10^3, 10^4, 10^5, 10^6\}$ triangularly distributed random variables, we obtain estimates and 95% asymptotic confidence intervals for $\mathbb{E}(M_n)$ for the cases $\theta \in \{0, 0.1, 0.2, 0.3, 0.4, 0.5\}$. The results are presented in Figure 3. Upon inspection of Figure 3, we note that $\mathbb{E}(M_n)$ remains close to 1.684567 for the assessed values of $n$. Further, for the cases where $\theta \in (0, 1)$, we observe that 2.369134 remains a conservative approximation, and that $\mathbb{E}(M_n)$ appears to converge to 2 for large $n$, as predicted by Huang and Shen (2007).

Table 2: Exact values of expected matches $\mathbb{E}(M_n)$ for various sample sizes $n$ and some supplementary values of the mode parameters $\theta$.

| $n\backslash\theta$ | 0.05 | 0.15 | 0.25 | 0.35 | 0.45 |
|---:|---:|---:|---:|---:|---:|
| 5 | 1.886329 | 2.033059 | 2.133066 | 2.188078 | 2.215196 |
| 10 | 2.107125 | 2.379993 | 2.529493 | 2.614819 | 2.654147 |
| 20 | 2.190642 | 2.522747 | 2.667543 | 2.73453 | 2.761107 |
| 50 | 2.32348 | 2.534738 | 2.511676 | 2.466411 | 2.439975 |
| 100 | 2.451241 | 2.417751 | 2.313203 | 2.261787 | 2.242005 |
| 200 | 2.458084 | 2.249248 | 2.181441 | 2.15757 | 2.148671 |
| 500 | 2.270841 | 2.127744 | 2.100175 | 2.089135 | 2.084782 |
| 1000 | 2.15722 | 2.083563 | 2.066782 | 2.059835 | 2.057063 |

## 5 Proofs of Theorems

### 5.1 Proof of Theorem 1

We can write

$$\mathbb{P}\left(\frac{j-1}{n} < X_{(j)} < \frac{j}{n}\right) = F_{j:n}\left(\frac{j}{n}\right) - F_{j:n}\left(\frac{j-1}{n}\right),$$

for each $j = 1, ..., n$. Thus, we can write

$$\begin{aligned}
\mathbb{E}(M_n) &= \sum_{j=1}^{n} \mathbb{P}\left(\frac{j-1}{n} < X_{(j)} < \frac{j}{n}\right) \\
&= \sum_{j=1}^{n} \left[F_{j:n}\left(\frac{j}{n}\right) - F_{j:n}\left(\frac{j-1}{n}\right)\right] \\
&= F_{n:n}(1) - F_{1:n}(0) + \sum_{j=1}^{n-1} a_j,
\end{aligned} \qquad (12)$$

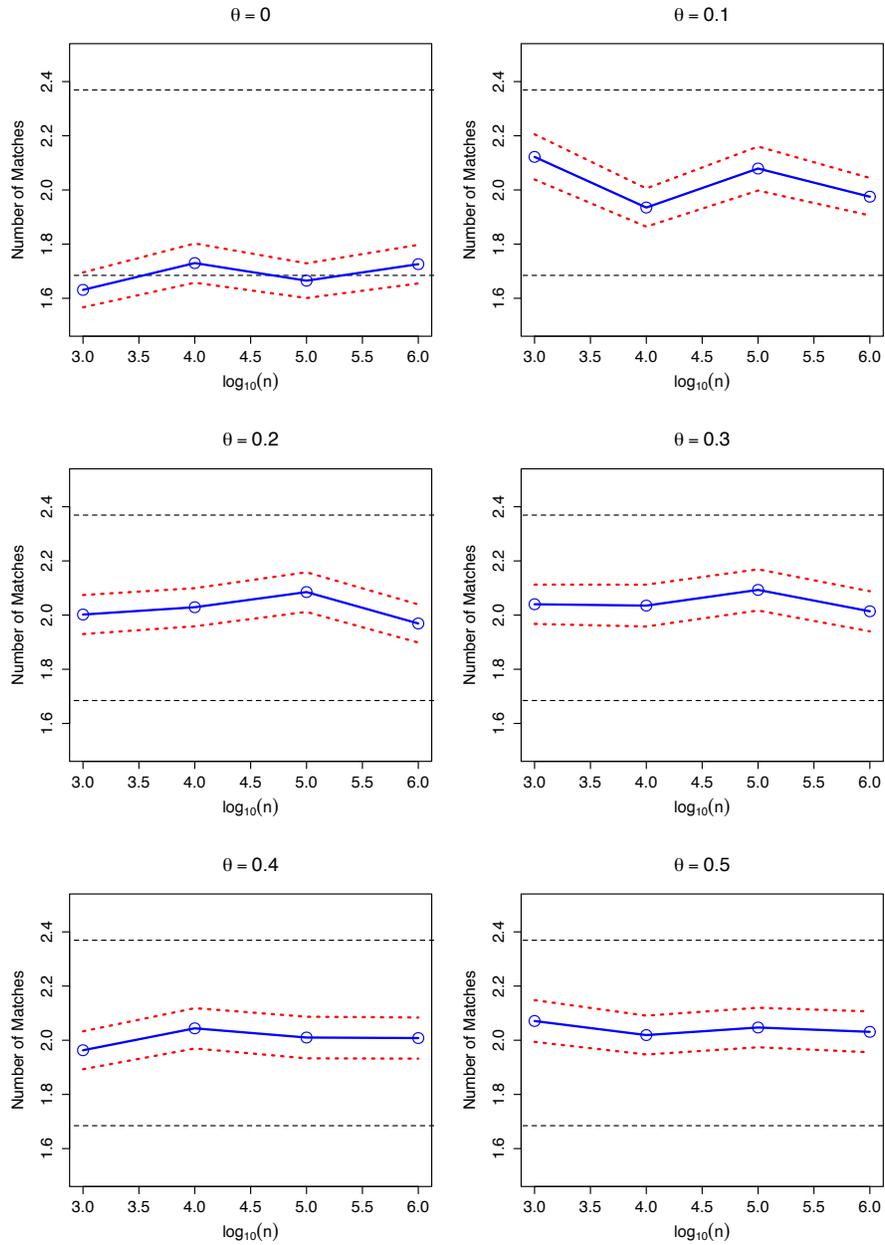

Figure 3: The ○ marked line indicates the average number of matches from 1000 Monte Carlo samples of $n$ observations. The dotted lines indicate the 95% asymptotic confidence interval for $\mathbb{E}(M_n)$. The lower and upper dashed lines mark the values 1.684567 and 2.369134, respectively. Each subplot presents results for a different value of $\theta$.

where $a_j = F_{j:n}(j/n) - F_{(j+1):n}(j/n)$. Using (6), we can write

$$a_j = \sum_{i=j}^{n} \binom{n}{i} F^i(x)[1-F(x)]^{n-i} - \sum_{i=j+1}^{n} \binom{n}{i} F^i(x)[1-F(x)]^{n-i}$$

$$= \binom{n}{j} F^j(x)[1-F(x)]^{n-j},$$

upon expansion of the summations. Finally, since $F_{1:n}(0) = 0$ and $F_{n:n}(1) = 1$ by definition of CDFs, we have the desired result by simplification of (12).

## 5.2 Proof of Corollary 1

Write (7) as

$$\mathbb{E}(M_n) = 1 + \sum_{j=1}^{n-1} \binom{n}{j} b_j(\phi_j), \tag{13}$$

where $b_j(\phi_j) = \phi_j^j (1-\phi_j)^{n-j}$ and $\phi_j = F(j/n)$. To obtain an upper bound for (7), we maximize (13) with respect to $\phi_j$, for $j = 1, ..., n-1$. Since each $\phi_j$ is linearly separable, we can maximize (13) by maximizing each $b_j$, respectively.

Solving the first-order condition using the derivatives

$$\frac{db_j}{d\phi_j} = (j - n\phi_j) \phi_j^{j-1} (1-\phi_j)^{n-j-1}$$

yields the solution $\phi_j^* = j/n$, for each $j$. For any $j$, $b_j$ is log-concave since it is the product of two powers of positive values [cf. Boyd and Vandenberghe (2004, Example 3.39)], and thus $b_j$ is also quasi-concave.

Note that $db_j/d\phi_j > 0$ when $\phi_j < \phi_j^*$ and $db_j/d\phi_j < 0$ when $\phi_j > \phi_j^*$. Thus $\phi_j^*$ is the mode and global maximizer of $b_j$ [cf. Boyd and Vandenberghe (2004, Sec. 3.4.2)]. Substitution of $\phi_j^*$ into (13) yields an upper bound for $\mathbb{E}(M_n)$. We obtain the desired result via approximation (5).

## 5.3 Proof of Lemma 1

Note that $s_j > 0$ for all $j = 1, ..., n$, and that the ratio $s_{j+1}/s_j = 2(1 + 1/j)^j / e^2 \to 2/e$ as $j \to \infty$. Since $2/e < 1$, we obtain the convergence of $S_n$ as $n \to \infty$, by the ratio test [cf. Khuri (2003, Thm. 5.2.6)]. The approximation $S_\infty \approx 0.684567$ is obtained via a partial sum of $n = 100$ terms.

## 5.4 Proof of Lemma 2

Note that $u_j^q > 0$ for all $j = 1, ..., n$, and $q > 0$. Consider the ratio $u_j^q/s_j = 2^q j^q j!/(j+q)! \to 2^q$ as $j \to \infty$. When $q$ is constant, we obtain convergence of $U_n^q$, as $n \to \infty$, since $S_n$ converges by the limit comparison test [cf. Khuri (2003, Thm. 5.2.5)]. The convergence of $V_j^p$ follows from the same argument.

# Acknowledgments

We thank the two anonymous reviewers for their useful comments, which have significantly improved the exposition of the article.

Cormen, T. H., Leiserson, C. E., Rivest, R. L., Stein, C., 2002. Introduction To Algorithms. MIT Press, Cambridge.

David, H. A., Nagaraja, H. N., 2003. Order Statistics. Wiley, New York.

Doane, D. P., 2004. Using simulation to teach distributions. Journal of Statistical Education 12, 1–21.

Glickman, T. S., Xu, F., 2008. The distribution of the product of two triangular random variables. Statistics and Probability Letters 78, 2821–2826.

Gunduz, S., Genc, A. I., 2015. The distribution of the quotient of two triangularly distributed random variables. Statistical Papers 56, 291–310.

Huang, J. S., Shen, P. S., 2007. More maximum likelihood oddities. Journal of Statistical Planning and Inference 137, 2151–2155.

Hunter, D. R., Lange, K., 2004. A tutorial on MM algorithms. The American Statistician 58, 30–37.

Karlis, D., Xekalaki, E., 2008. The polygonal distribution. In: Minguez, R., Sarabia, J.-M., Balakrishnan, N., Arnold, B. C. (Eds.), Advances in Mathematical and Statistical Modeling. Birkhauser, Boston, pp. 21–33.

Khuri, A. L., 2003. Advanced Calculus with Applications in Statistics. Wiley, New York.

Kotz, S., Van Dorp, J. R., 2004. Beyond Beta: Other Continuous Families of Distributions with Bounded Support and Applications. World Scientific, Singapore.

Lucas, A., Scholz, I., Boehme, R., Jasson, S., Maechler, M., 2014. gmp: Multiple Precision Arithmetic.
URL https://CRAN.R-project.org/package=gmp
18